\documentclass[prd,letterpaper,superscriptaddress,reprint,nofootinbib]{elsarticle}

\usepackage{hhline}

\usepackage{natbib}
\begin{document}
\begin{frontmatter}
\title{\boldmath Polarization Angle  Dependence of Vertically Propagating Radio-Frequency Signals in South Polar Ice} 
\author[ku]{DaveZ Besson}
\author[unl]{Ilya Kravchenko}
\author[iit]{Krishna Nivedita}

\address[ku]{Dept. of Physics and Astronomy, Univ. of Kansas, Lawrence, KS, 66044, USA}
\address[unl]{Dept. of Physics and Astronomy, Univ. of Nebraska-Lincoln, NE, 68588, USA}
\address[iit]{Department of astrophysics / IMAPP, Radboud University,  P.O Box 9010, 6500 GL, Nijmegen, The Netherlands}
\date{\today}
\begin{abstract}
To better understand the effect of ice properties on the science reach of radio experiments designed to measure ultra-high energy neutrinos (UHEN), we recently considered the timing and amplitude characteristics of radio-frequency (RF) signals propagating along multi-kilometer, primarily horizontal trajectories through cold Polar ice at the South Pole. That analysis indicated satisfactory agreement with a model of ice birefringence based on ice crystal (${\hat c}$-axis) data culled from the South Pole Ice Core Experiment (SPICE). Here we explore the geometrically complementary case of signals propagating along primarily vertical trajectories, using extant data from the Askaryan Radio Array (ARA) experiment, supplemented by a refined analysis of older RICE experimental data. The timing characteristics of the South Polar data are in general agreement with the same birefringence model, although a several nanosecond discrepancy is found in comparison to Taylor Dome data. Re-analysis of older RICE data also confirm the correlation of signal amplitudes reflected from internal-layers with the direction of ice flow, similar to previous observations made along a traverse from Dome Fuji to the Antarctic coast. These results have two important implications for radio-based UHEN experiments: i) if birefringence can be locally calibrated, the timing characteristics of signals propagating from neutrino-ice interactions to a distant receiver might be used to infer the distance-to-vertex, which is necessary to estimate the energy of the progenitor neutrino, ii) the measured reflectivity of internal layers may result in previously-unanticipated backgrounds to UHEN searches, requiring significantly more modeling and analysis.
\end{abstract}
\end{frontmatter}

\message{Note that propagation velocity is defined by polarization, not k, separately for p (perpendicular/parallel to plane of incidence) or
s (in plane of incidence)=>explains 50 ns shift observed for SPQR09 (DUH - OF COURSE THIS IS A POLARIZATION-DEPENDENT EFFECT, AS THE POLARIZATION DETERMINES THE DIRECTION OF THE PERTURBATION TO THE ATOMIC ELECTRONS TRANSPORTING THE SIGNAL!)}

\section{Introduction}
The field of radio-frequency neutrino detection via the Askaryan Effect\citep{Askaryan1962a,Askaryan1965} has undergone steady, and significant growth since its experimental inception approximately one-quarter century ago. The early 1990's marked the first detailed simulation of the coherent microwave signal expected for in-ice neutrino interactions\citep{ZasHalzenStanev1992}, as well as a full-scale experimental simulation\citep{Frichter:1995cn}. Deployment of hardware\citep{KravchenkoFrichterSeckel2003,Kravchenko:2006qc} and test beam experiments\citep{SaltzbergGorhamWalz2001,AskaryanSLAC2} verifying the expected signal followed soon thereafter. The polar ice sheets, offering a large, stable, and largely homogeneous radio-transparent\citep{barwick2005south} neutrino target comprise an attractive target medium. 

Such experiments have pursued three different deployment strategies. The ARIANNA experiment, initially deployed at Moore's Bay, Antarctica, and later also at the South Pole\citep{Barwick:2014pca,Barwick:2014rca,Barwick:2016mxm}, features high-gain antennas deployed on the surface, scanning primarily downwards into the ice for upcoming signals, but also with excellent sensitivity to down-coming radio signals from air showers induced by cosmic rays\citep{Barwick:2016mxm}. This approach offers ease and speed of deployment (with a full, 8-channel station installed and commissioned in a single day), a monolithic antenna design that minimizes systematic errors as well as cable runs, and accessibility of hardware for repair. In-ice antenna deployment circumvents antenna response uncertainties associated with the air/ice boundary layer, but requires separate antennas tailored to primarily vertical vs. horizontal polarization, and, quite significantly, also requires ice borehole drilling. The deep hole deployment strategy was employed by the original RICE\citep{Kravchenko:2011im,kravchenko2004situ,Kravchenko:2003gk,KravchenkoFrichterSeckel2003,KravchenkoFrichterSeckel2003} and the successor ARA\citep{allison2012design,allison2019design,allison2020constraints} experiments. The Radio Neutrino Observatory in Greenland (RNO-G), which began operation at Summit Station, Greenland in the summer of 2021\citep{Aguilar:2020xnc} combines both the surface, as well as the in-ice deployment strategy to simultaneously maximize sensitivity to both down-coming cosmic rays, as well as upcoming neutrino signals.
Alternately, a suite of radio receivers tethered to a high-altitude (38 km) balloon, as with the ANITA\citep{GorhamAllisonBarwick2009} and PUEO\citep{abarr2021payload} experiments. can scan an enormous volume of ice with high-gain horn antennas, albeit at the cost of an elevated experimental neutrino energy threshold (owing to the large typical distance to neutrino interaction point) and the logistical overhead, as well as the reduced livetime, associated with balloon-borne payloads. Nevertheless, this elevated strategy also yielded the first results on cosmic rays detected via self-triggering of a radio array\citep{HooverNamGorham2010}. Recently, an alternate strategy has emerged, based on surveying ice for echoes from an in-ice radar transmitter\citep{Prohira:2019glh,prohira2021radar}, as with the Radar Echo Telescope (RET), currently slated for polar deployment beginning in 2024.

It has long been realized that the sensitivity of such experiments is determined, in large part, by the dielectric properties, particularly refraction, attenuation, and polarization-dependent asymmetries of the polar ice itself. Specifically,
signal identification for such experiments is complicated by several glaciological effects, as follows:
\begin{itemize}
\item Both modeling and experiment, for example, have indicated unexpected radio-frequency transport and propagation modes\citep{barwick2018observation,deaconu2018measurements,telescope2021modeling} attributed to density fluctuations in the ice medium, which may impact the scientific mission of planned neutrino observatories. Short-duration (nanosecond-scale) radio-frequency signals propagating over the $\sim$km-scale baselines typical of neutrinos may exhibit measurable variations in amplitude over modest differences in reception geometry, otherwise unexpected for a uniform, isotropic target. A rough estimate of the scale of ice density perturbations required to yield observable effects can be obtained from simple arguments. In a ray optics picture, two one-meter wavelength (f=300 MHz) signals traveling in a medium with density (and therefore refractive index) fluctuations of order 1.5 parts per mille shift by $\delta\phi=\pi$/2 for each 0.25 km of pathlength. In the conventional H\"uygens wave optics picture, such density fluctuations can disrupt the detailed cancellation of wavelets that would otherwise lead to `shadow' zones, resulting in some signal observed from otherwise classically-forbidden source regions (as measured in data)\citep{barwick2018observation}. 
\item Birefringence will lead to not only a variation of signal velocity with signal polarization, but can also lead to signal `splitting', when projected onto the basis states of the crystal-orientation tensor. Construction, and testing of a model that can predict birefringent effects at any given site is therefore essential to a sound estimate of the experimental sensitivity of any present, or planned neutrino experiment. 
\item Layered impurities within the ice sheet can comprise internal reflectors that disrupt `standard' rectilinear ray trajectory reconstruction. Moreover, such internal layering may lead to an `upcoming' ray bundle, apparently originating within the ice sheet itself, and therefore likely to evade the surface background suppression strategies typical of extant neutrino experiments. For a given wavelength, internal layers with spacing comparable to that wavelength may also effect signal channeling phenomena that can also amplify potential backgrounds.
\end{itemize}
In this paper, we address the last two of these points. We first discuss birefringence phenomenology, then confront experimental data drawn from the South Pole (taken with the RICE and ARA experiments) as well as data taken at Taylor Dome, Antarctica. For the RICE experiment, polarization-dependent propagation time asymmetries are quantified, as a function of depth into the icesheet, by measuring the total transit time for an RF signal sent from the surface, reflected from an internal layer, and subsequently measured at the surface. In particular, we search for correlations with the local ice flow direction, which, with gravity, constitute the two internal stresses within the ice sheet. As an ancillary measurement, we consider the reflection strength for the observed internal layers; interestingly, we also measure a reflection amplitude dependence on polarization orientation for all internal layers considered, as documented below. 

\section{Polar Ice Properties}
\subsection{Internal Layers}
Previous radar probes of the Antarctic ice sheet have demonstrated the presence of radio-frequency reflectors, internal to the ice sheet\citep{matsuoka2003crystal,fujita2003scattering,fujita2006radio,Kenny2012Fabric}. These may be associated with discrete conducting layers, resulting, e.g., from volcanic ash deposition, or regions within the ice where the crystal fabric changes its structure. For the purposes of this analysis, discrete layers provide convenient internal reflectors, allowing us to sample a wide range of depths and echo return times.

\subsection{Birefringence}
\begin{figure}[htpb]
\includegraphics[width=\textwidth]{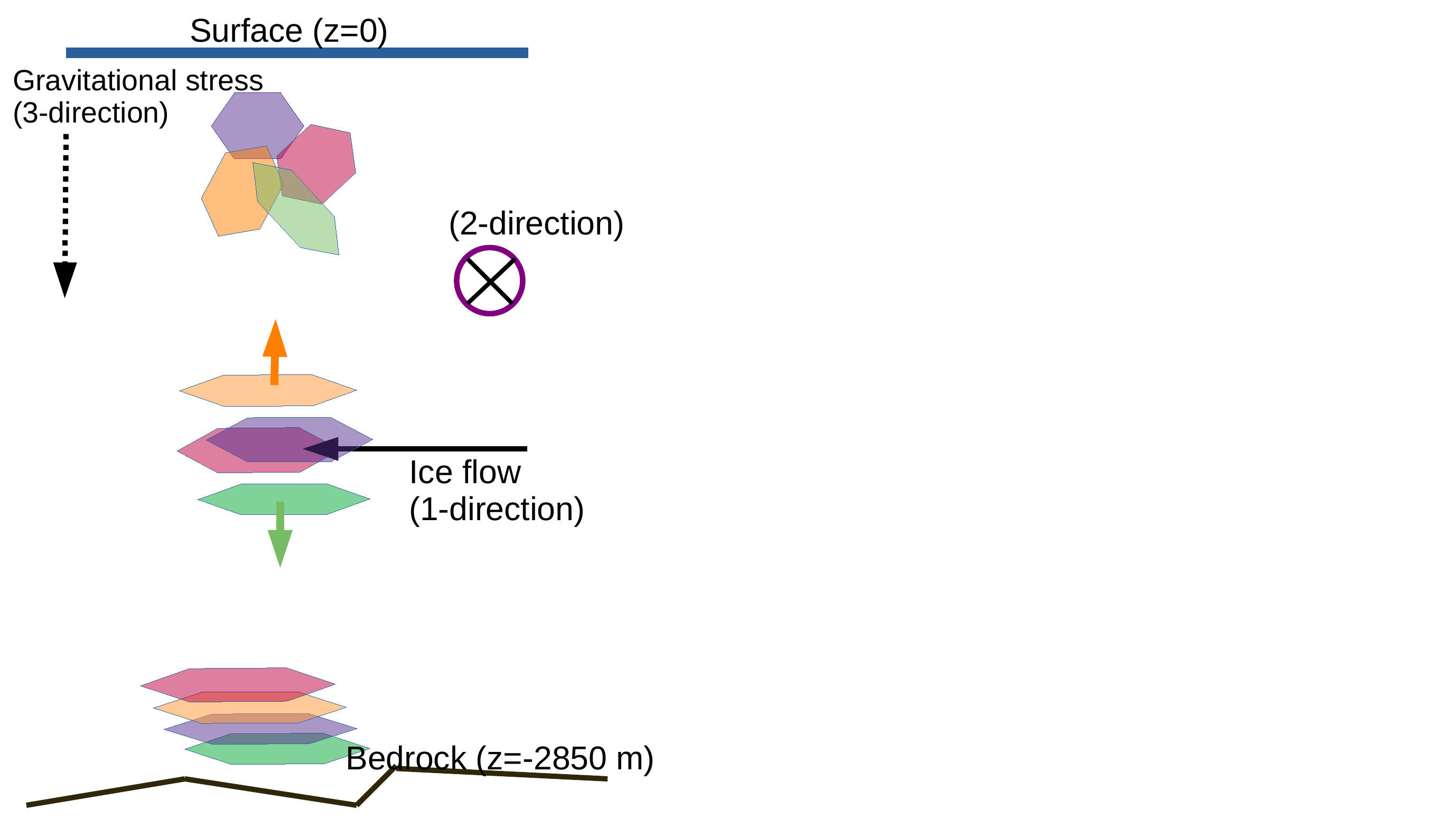} 
\caption{Illustration of evolution of crystal-orientation fabric (COF), as a function of depth into the ice sheet. At shallow depths, where horizontal strain and vertical gravitational stress acting on more recently deposited snow are mild, hexagonal ice crystals are typically randomly oriented. At intermediate depths, which sample more of the ice sheet strain (and flow) history, crystals elongate in the direction of flow; the c-axis distribution is perpendicular to the crystal face and forms a `girdle'. Close to the bedrock, gravitational overburden dominates, and the ${\hat c}$-axis distribution prefers a uniaxial alignment with the z-axis.}
\label{fig:cof}
\end{figure}

Absent external stresses, isolated ice crystals have roughly six-fold planar rotational symmetry\citep{bragg1921crystal}. Within the ice sheet, ice crystals are subject to strain along both vertical (due to the gravitational overburden) and also horizontal (due to ice flow) directions\citep{rigsby1960crystal,alley1995mapping}. In the `standard' picture, the ${\hat c}$-axis distribution is expected to be approximately isotropic, with no preferred spatial direction, for shallow ice with modest overburdens. At intermediate depths, orientations for which the plane of the ice crystal is perpendicular to the ice flow direction are disfavored; the geometric distribution of vectors normal to the ice crystal plane is correspondingly described as a `girdle', with ice crystals slightly elongated along the ice flow and the ${\hat c}$-axes correspondingly distributed in a plane transverse to the flow direction. At depths closest to the bedrock, the gravitational overburden forces the ${\hat c}$-axis to increasingly align with ${\hat z}$- and favors a uniaxial distribution. The cartoon shown in Figure \ref{fig:cof} illustrates the evolution of the crystal-orientation-fabric with depth.

Of particular interest here is the relationship between the measured ${\hat c}$-axis distribution and the corresponding expectations for birefringent effects. In our comparison of birefringence data to expectation, we follow the biaxial model developed by Jordan {\it et al.}\citep{jordan2020modeling}, based on the South Pole Ice Core Experiment (SPICE) thin slice ice fabric measurements\citep{Voigt2017} and data on light velocity through a crystal as a function of light polarization axis, derived from Fujita {\it et al.}\citep{fujita2003scattering,matsuoka1997precise}. Following \citep{fujita2003scattering}, we designate an orthogonal coordinate system onto which the signal polarization is projected, with basis vectors defined as: the ``1-'' polarization direction parallel to the ice-flow direction in the horizontal plane, the ``3-'' direction parallel to z-, and the ``2-'' direction perpendicular to the ice-flow direction in the horizontal plane and orthogonal to both ``1-'' and ``3-''. Ice fabric measurements from the SPICE core quantify the degree of alignment of the ${\hat c}$-axis with these three directions. 
For uniaxial laboratory ice, the asymmetry in the refractive index between the 3- direction and the orthogonal 1- or 2- directions is measured to be $\sim$0.018\citep{matsuoka1997precise}. For the case of signal propagation in the z-direction (e.g. as for radar echoes from the Antarctic surface, through the ice, and reflected from the bedrock), the polarization axis lies in the horizontal (1-2) plane.  As the signal propagates through the ice, an instantaneous, local spherical coordinate frame is defined by the s- and p- polarization vectors, with s- aligned with the local ${\hat\phi}$ azimuthal direction (consistent with the standard definition in spherical coordinates) and p- aligned with the local ${\hat\theta}$ direction (defined by ${\hat k}\times{\hat\phi}$ at any point along the ray trajectory, such that ${\hat\theta}$ is, in general, inclined relative to the vertical [z-] axis). 
In this model, for purely horizontal k-vectors, SPICE ${\hat c}$-axis measurements imply that signal polarizations perpendicular to the ice flow direction, and aligned with the 2- axis 
 will temporally lag behind signal polarizations parallel to the ice flow direction. 

\section{Experimental Data}
\subsection{Previous Birefringence Measurements from the Askaryan Radio Array (ARA)}
\subsubsection{Horizontal propagation data using the SPICE core transmitter calibration}
In 2018-19, a `fat' dipole transmitter was lowered into the ice hole drilled for the South Pole Ice Core Experiment (SPICE); the five Askaryan Radio Array (ARA) receiver stations, ranging from 2--5 km distant and with receivers at 50--200 m depths, recorded signal waveforms\citep{allison2020long}. As reported previously, the measured time asymmetry for propagation along predominantly horizontal trajectories, between polarizations primarily along z- vs. polarizations in the horizontal plane, is qualitatively consistent with the Jordan {\it et al.} model\citep{jordan2020modeling}. For propagation directions parallel to the ice flow, the horizontally propagating ${\vec E}$-field vector is almost entirely along the 2- direction and the polarization lies in the (2-3) plane; since $n_2\approx n_3$, no time delay is expected between signal arrival times for vertical vs. horizontal polarizations, consistent with observation. For propagation directions perpendicular to the ice flow, the horizontally propagating ${\vec E}$-field vector is almost entirely along the 1- direction, resulting in the VPol signal lagging the HPol signal by some tens of nanoseconds for these baselines, since $n_1<n_3$. 

Two effects slightly complicate interpretation of those data: First, as the transmitter antenna is lowered into the SPICE core icehole, the p-polarization increasingly deviates from true vertical and acquires an increasing horizontal polarization projection component (in addition to the known $\sim$-10 dB cross-polarization power of the dipole transmitter antenna). Second, we experimentally often observe unexpectedly large power for horizontal polarizations. For the ARA01 station receiver array, for example, corresponding to the steepest incidence angles (approaching 30 degrees with respect to vertical), the measured HPol power has been observed to often exceed the VPol power\citep{allison2020long}; correcting for the more-limited bandwidth of the HPol antennas only exacerbates this excess. \message{Although unlikely to exceed $\sim$1 ns, frequency-dependent group delays are likely to differ, at least slightly, between VPol and HPol. Moreover, the pattern of observed HPol:VPol power, as a function of transmitter depth, is not readily accommodated by the expectation for a simple dipole beam pattern.}

Some of the unexpectedly large HPol power may be a direct consequence of birefringence and coupling between the s- and p-modes, as highlighted recently in more rigorous treatments than Jordan {\it et al.}\citep{Connolly:2021cum,heyer2022first}. For an isotropic, uniform medium, the local electromagnetic wavespeed is dependent only on the local refractive index profile, independent of both ${\hat k}$ and polarization ${\hat P}$. In a biaxial medium, wavespeed depends on the directional excitation of the ice crystal transporting the signal. For polarizations aligned with ${\hat 1-}$, ${\hat 2-}$ or ${\hat 3-}$, the net energy flux vector (Poynting vector ${\hat S}$) measured by an observer is parallel to ${\hat k}$. For off-axis signals with polarizations making angles ${\chi_1}$, ${\chi_2}$ and ${\chi_3}$ relative to the underlying basis vectors, the non-uniform molecular vibration of the medium `projects' an arbitrary polarization onto the underlying birefringence basis; the transversality condition (${\hat E}\times{\hat H}={\hat k}$) which follows from Maxwell's equations therefore implies that the allowed modes, in general, will have some angular deviation $\psi$ between ${\hat k}$ (given by ${\vec D}\times{\vec B}/{\vec H}$) and ${\hat S}$ (given by ${\vec E}\times{\vec B}/{\vec H}$). Geometrically, whereas in an isotropic medium, energy flows spherically outwards from a given source, in a birefringent medium, the direction of net energy flow expands along embedded ellipsoids, with an angular deviation from spherical flow proportional to the refractive index asymmetry. In the presence of possible interference\footnote{In principle, the transfer function of the receiver could be inverted to improve the temporal separation of the two basis state signals.} (and exacerbated by the cross-polarization response of a given antenna), the amplitude of a transmitted continuous wave signal at a receiver will therefore depend on the angles ${\chi_1}$, ${\chi_2}$, ${\chi_3}$ and the propagation distance. For short-duration signals, such that the temporal extent of signal $t_{signal}$ is much smaller than the combined effects of the intrinsic transmitter, receiver and system electronics response time and group delays $t_{response}$, signal may similarly interfere at the receiver, or in the data acquistion (DAQ) hardware. 

\subsubsection{Birefringence measurements from embedded transmitters}
In addition to the SPICE core transmitter, the so-called `deep transmitters' co-deployed with IceCube photomultiplier modules on IceCube strings 1 and 22 in 2011 also provide a `standard' calibration source, as reported elsewhere\citep{allison2019measurement,allison2012design}. Two shallower transmitters (IC1S and IC22S) were deployed at a depth of 1400 meters; one transmitter was deployed at a depth of 2400 meters in IceCube hole 1 (IC1D). Ray trajectories to the so-called ARA `testbed' receiver station from IC1D are inclined at nearly 45 degrees relative to the vertical; projected onto the horizontal plane, a ray path from the deep pulsers to the testbed makes an azimuthal angle of approximately 30 degrees with respect to the local horizontal ice flow direction. Unlike the ARA stations deployed subsequent to the testbed, for which antennas were configured into eight HPol/VPol pairs vertically offset by $\sim$2 m, the placement of the testbed antennas was less standardized. Given the deployment geometry, VPol antennas are primarily sensitive to polarizations parallel to the $n_3$ axis, whereas HPol antennas at the testbed site are primarily sensitive to the $n_2$ axis; the arrival time difference of propagating signals therefore provides a measure of the difference $n_3-n_2$. Into each of four testbed ice boreholes, one HPol and one VPol antenna were deployed, with vertical separations between H and V varying from --6$\to$+6 meters. Numerically, however, we can take advantage of the fact that the summed depth of the four antennas of each polarization are approximately equal (to within 0.5 m), so summed propagation time differences between HPol/VPol pairs should (modulo refractive index biases) largely cancel; i.e., in the absence of birefringence $\Sigma_i(t(arrival_i^{Hpol})\approx \Sigma_i(t(arrival_i^{Vpol})$. Calculating these sums for approximately 1000 transmitter events per pulser collected during January, 2011, we find a per-channel HPol-VPol arrival time separation of $\sim$7-9 ns for all three of the pulsers\citep{allison2020long}. For each pulser's 1000-event sample, the distribution of calculated testbed H-V time differences has a width of approximately 2 nanoseconds, as indicated in the Table below.\footnote{Differing by several nanoseconds, these results are obtained using the same raw data as previously analyzed, albeit with slightly different software requirements, yielding a higher precision and higher purity sample. Previous results were based on a simple voltage threshold-crossing criterion to extract hit times, and therefore time-differences between channels. Revised results are based on cross-correlation of the Hilbert envelope of waveforms, which uses the full frequency band information, are more robust to slewing effects, and has been demonstrated to have higher efficiency and also give more precise signal source locations, based on embedded calibration transmitter pulser reconstruction.} 

\subsection{Data taken with the RICE experiment}
\begin{figure}[htpb]\centering    
\includegraphics[height = 0.9 \textwidth, width = \textwidth]{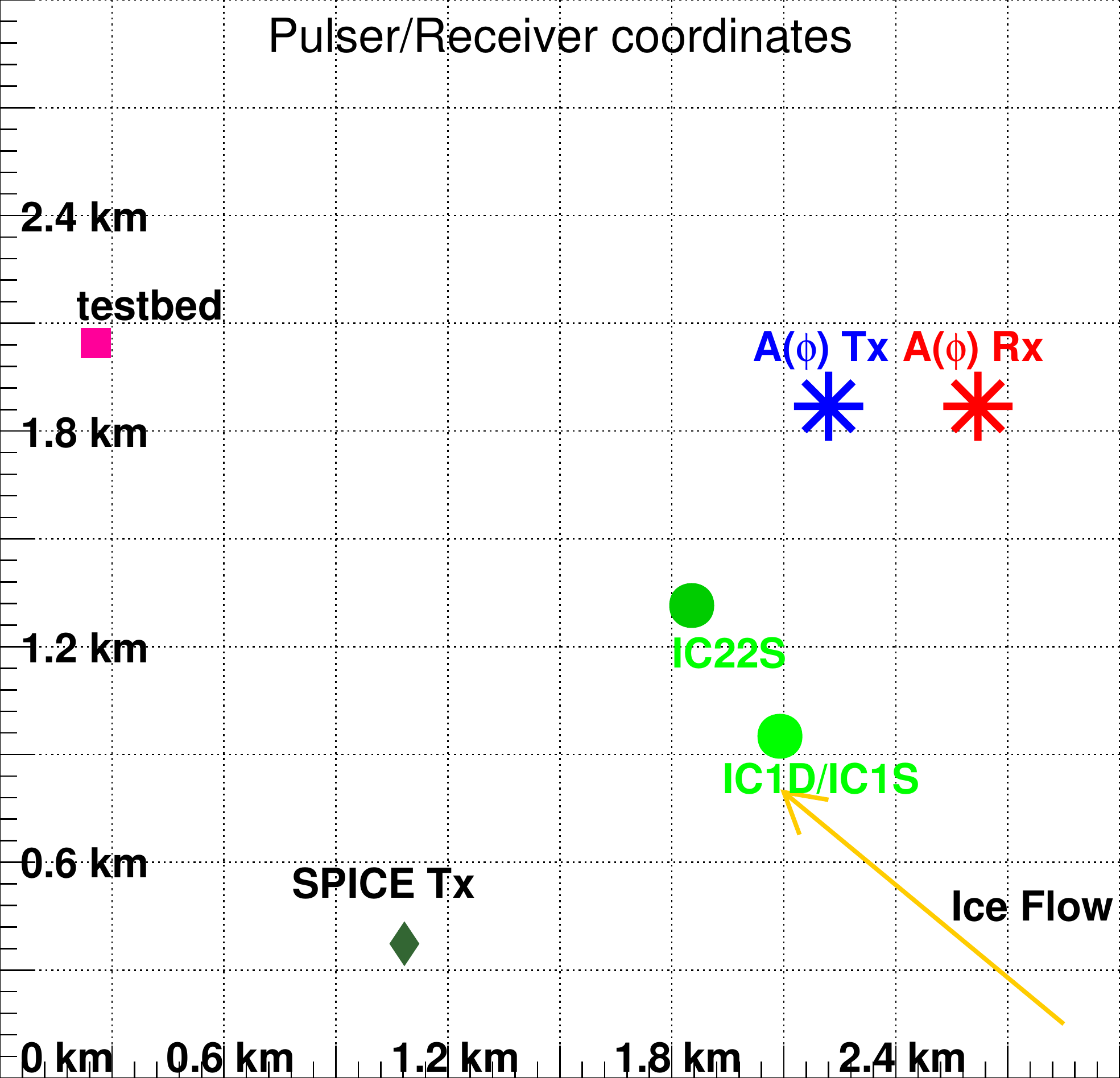}    
\caption{Map illustrating relative positions/layout of measurements made at South Pole, including bedrock and internal layer measurements using RICE experimental hardware (shown as asterisks, with Amplitude measured as a function of azimuth $\phi$) and broadcasts from the in-ice deep pulser IC1D, and shallower pulsers IC1S and IC22S, to the near-surface testbed radio receiver station.}    
\label{fig:geo}
\end{figure}

As the first experiment of its kind in the polar regions, one of the primary RICE science goals was to provide a detailed {\it in situ} characterization of the neutrino ice target. In addition to the twenty radio receivers that comprised the primary neutrino detector, RICE hardware also included ancillary antennas, signal generators, and digital oscilloscope data acquisition for remote measurements in the field.
For bistatic radar data taken in 2007 by the RICE experiment and detailed below, two horn antennas, separated by $\sim$100-m, were placed on the ice surface approximately 1 km from the geographic South Pole, with the horn faces directed downwards into the snow\citep{Besson:2009zza,Besson:2010ww}. A map showing the location of the RICE data-taking, superimposed on the local geometry of the ARA testbed and also the IceCube deep pulsers, is shown in Fig. \ref{fig:geo}.

As the two horn antennas were co-rotated and ns-duration signals fed into the transmitter horn antenna, echoes were recorded in the receiver horn antenna, in principle resulting from both discrete internal layer as well as continuous volumetric signal scattering. The goal of that original analysis was to measure birefringence, as a function of depth, to similarly infer the depth dependence of ice properties. Those RICE data indicated simultaneity of echo returns, for all azimuthal orientations, up to 13 microsecond return times, although the analysis at that time still allowed 5--10 ns systematic errors. 
Although not systematically quantified, those data also suggested an unexpectedly large variation in signal return strength, as a function of azimuthal angle $\phi$ relative to the local ice flow direction (approximately 40 degrees west of grid North).  A nearly identical study at Taylor Dome revealed a measurable, $\sim$10-nanosecond time shift, relative to a total travel time-to-bedrock of $\sim$11700 ns, between the fast/slow azimuthal orientations, observed to be perpendicular to each other (Ref. \citep{besson2008situ}, 
for which the cross-correlation is maximized at a time delay of 6.8 nanoseconds).  
However, the Taylor Dome crystal-orientation fabric indicate only a very small difference in the $n_1$ and $n_2$ eigenvalues to a depth of 300 meters, and transition to uniaxial fabric at greater depths\citep{kennedy2013evolution}. 

Herein, we have conducted a reanalysis of the earlier RICE data using improved signal processing software, particularly for identifying radar echoes from embedded internal layers. For the 2007 RICE data sample, we identify internal layer echoes by their observed (uncorrected for intrinsic cable delays, measured to be $\sim$190 ns) return times, corresponding to a bright echo at 6 $\mu$s return time, a triplet of echoes between 9.7 and 10 microseconds, a second triplet of echoes between 13.7 and 14 microseconds, and weaker echoes at 17.3 and 19.6 microseconds, respectively.  Whereas the previous analysis relied on simple `by-eye' estimates of signal timing and amplitudes, our newer analysis employs a complete suite of radio-frequency processing software, developed over the last two decades for use by the ARA and ANITA experiments. Threshold-crossing times for the Hilbert-transformed voltage(time) traces afford more precise estimates of signal timing. In cases where the difference in threshold-crossing time for two waveforms $\delta$t, corresponding to orthogonal polarizations, e.g., are compared, the rms of a cross-correlation graph between the two waveforms allows an estimate of our statistical uncertainty for a given measurement. The systematic error inherent in our measurements can be calculated by varying the software passband of the system by $\pm$100 MHz.  

\subsubsection{Signal wave speed dependence on polarization angle}
\begin{figure}[htpb]\centerline{\includegraphics[width=0.8\textwidth]{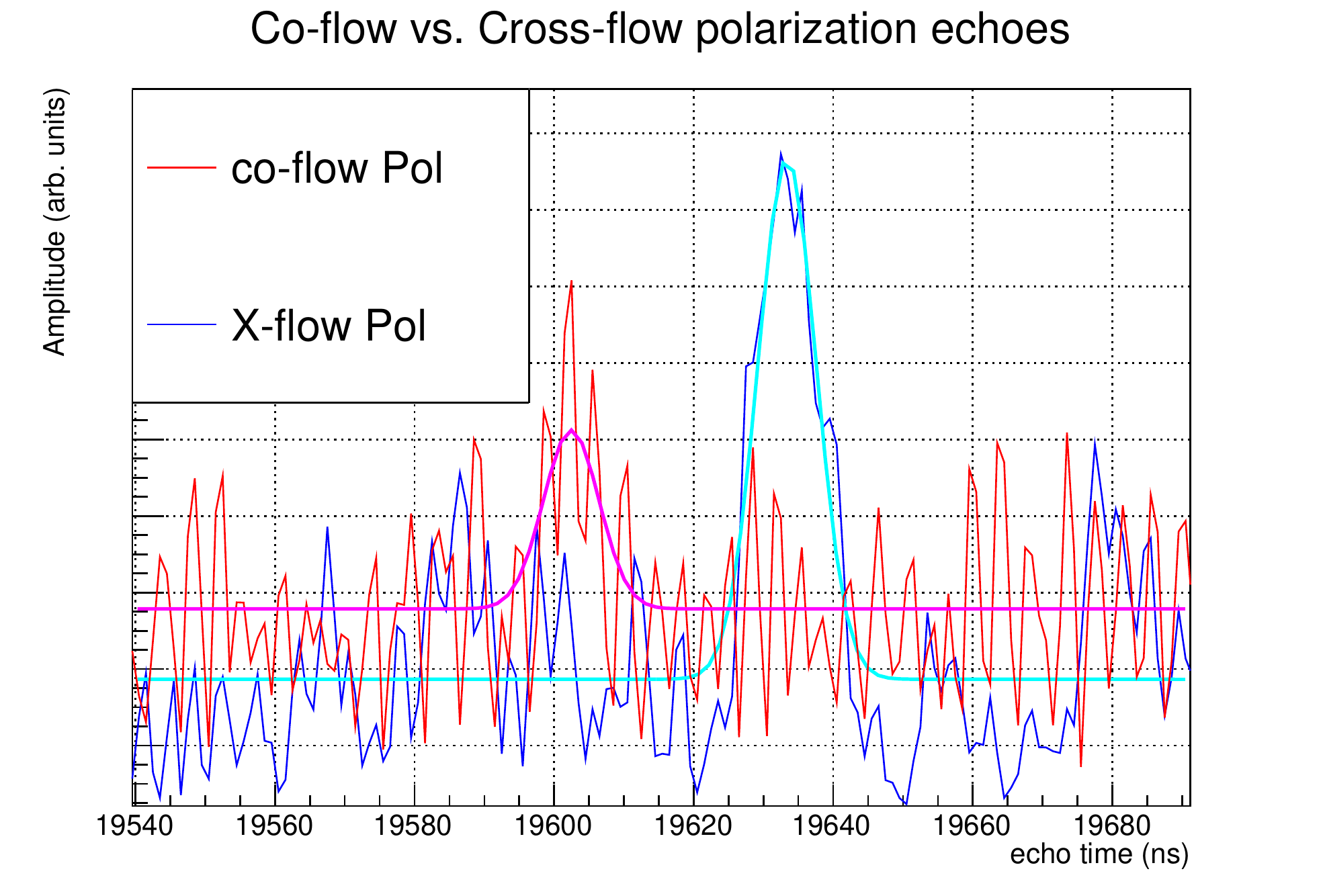}}\caption{Hilbert Transform of RICE bi-static radar echo voltage traces, measured around 19.6 $\mu$s return time. Overlaid are Gaussian fits to most significant excursions for polarizations aligned either along- or transverse to local flow direction, as indicated in Figure.}\label{fig:19635.pdf}\end{figure}

The greatest fractional temporal resolution was obtained from the deepest observed internal layer echo at 19.6 microseconds. Figure \ref{fig:19635.pdf} shows the (normalized) Hilbert transform of the recorded voltage traces, for azimuthal polarizations either parallel or perpendicular to ice flow, with fits to a Gaussian signal plus a first-order polynomial function overlaid. Our revised analysis indicates a temporal shift for polarizations parallel to the ice flow direction relative to polarizations perpendicular to the ice flow direction.  Combined with similar measurements for the earlier echoes, as well as bedrock echoes and measurements made at Taylor Dome, Table \ref{tab:comp} compares 
our propagation time data points with model predictions from Jordan {\it et al.} Although the Jordan {\it et al.} model qualitatively reproduces some the features observed in our South Pole data\citep{jordan2020modeling}, the Taylor Dome data deviate slightly from model prediction, at the level of several nanoseconds. We remind that a more complete birefringence model, including inter-modal coupling has been implemented for both a fixed frequency of 300 MHz\citep{Connolly:2021cum} and also the case of broadcast pulses, more applicable to this analysis\citep{heyer2022first}. Those calculations demonstrate that mode-mixing and also possible interference effects may result in significant suppression/enhancement of signal polarizations other than those originally broadcast at a transmitter, although time delay predictions are generally consistent with the Jordan {\it et al.} model. 
\begin{table}[htpb]
\centering
\begin{tabular}{cccccc} \\ \hline
Expt & Measurement & Geometry & Data $\delta$t& Model $\delta$t & Measures \\ \hline
RICE & 6 $\mu$s echo & Bistatic Radar & -2.1$\pm$2.6 ns & -3.2 ns & $n_2-n_1$ \\
RICE & 9.8 $\mu$s echo & Bistatic Radar & -15.7$\pm$7 ns & -8.4 ns & $n_2-n_1$   \\
RICE & 19.6 $\mu$s echo & Bistatic Radar & -24$\pm$4.5 ns & -28.8 ns & $n_2-n_1$   \\
RICE & 34 $\mu$s echo & Bistatic Radar & -55$\pm$3.2 ns & -67.3 ns & $n_2-n_1$  \\
Testbed & IC1S$\to$testbed & deep pulser & -7.3$\pm$1.8 ns & -5.2 ns & $n_2-n_1$  \\
Testbed & IC22S$\to$testbed & deep pulser & -7.9$\pm$1.8 ns & -5.2 ns & $n_2-n_1$  \\
Testbed & IC1D$\to$testbed & deep pulser & -8.6$\pm$1.9 ns & -5.7 ns & $n_3-n_2$  \\
Taylor Dome & 11.7 $\mu$s bed echo & Bistatic Radar & -6.8$\pm$2.1 ns & -1.3 ns & $n_2-n_1$  \\ \hline
\end{tabular}
\caption{Comparison of data to birefringence model predictions. South Pole prediction\citep{jordan2020modeling} is based on SPICE ice fabric measurements\citep{Voigt2017} to a depth of 1700 meters; prediction for $z<-1700$ m assumes a linear extrapolation to the bedrock. A more realistic extrapolation might take into account the expected transition to a single-pole uniaxial c-axis orientation in the lower third of the ice sheet (and below the maximum depth for which SPICE core data are available). The maximum Taylor Dome $\delta$t data value was determined empirically, although no polarization orientation was recorded relative to the local ice flow at the time the radar data were taken. Taylor Dome COF data\citep{kennedy2013evolution} indicate minimal expected time birefringent asymmetry, to a depth of 500 meters. Shown are combined statistical (based on the width of the cross-correlation graph) plus systematic (based on varying the experimental frequency passband) errors, with the former typically exceeding the latter.}\label{tab:comp}
\end{table}


\subsubsection{Amplitude dependence on polarization angle}
\begin{figure}[htpb]\centerline{\includegraphics[width=0.8\textwidth]{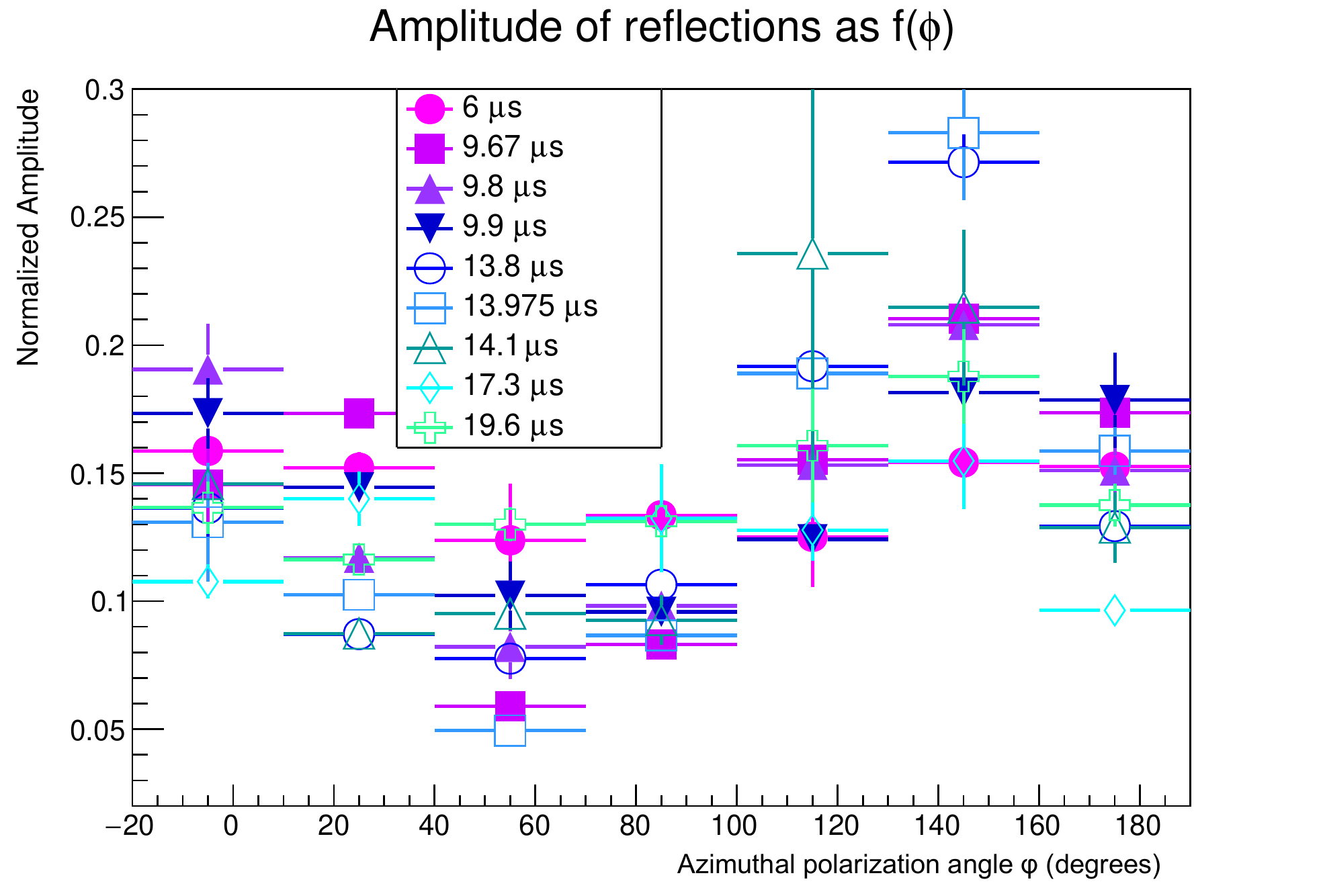}
}\caption{Peak measured amplitude dependence of internal layer echoes on azimuthal signal polarization angle $\phi$, as measured in horizontal plane.}\label{fig:ASigVphi}\end{figure}

Figure \ref{fig:ASigVphi} illustrates the correlation of echo strength with azimuthal signal polarization angle; we use a local coordinate system for which the ice flow direction corresponds to $\phi\approx 60^\circ$. 
As a sample calculation, for the open light blue squares in the Figure, the ratio of the peak amplitude at 150 degrees to the minimum amplitude at 60 degrees is a factor of 4.6 in field, or 13.3 dB in power.
The minimum power variation is observed for the 6 microsecond return, corresponding to a ratio of only 3 dB in power. Our data indicate no systematic trend with echo depth. Averaging over the calculated data points, we quantify the average ratio of 150 degree to 60 degree power to be 8.1$\pm$3.2 dB.
Measurements of signal returns along a traverse from Dome Fuji to the Antarctic coast also revealed a strong dependence of reflected signal amplitude on the signal polarization angle, referenced to the local ice flow direction; the echo return power for polarizations aligned with the local ice flow direction were observed to be $\sim$10 dB smaller than for polarizations perpendicular to the local ice flow direction\citep{matsuoka2003crystal}. Our observations are consistent with that result, as well as a preliminary version of this analysis\citep{dzbilya2013cryo}.
We note that the Dome Fuji traverse also reported observable birefringent effects (using a different, interference-based analysis technique compared to that used in this analysis) throughout the upper 1 km of the ice sheet, consistent with the values reported here.

\begin{figure}[htpb]\centerline{\includegraphics[width=0.8\textwidth]{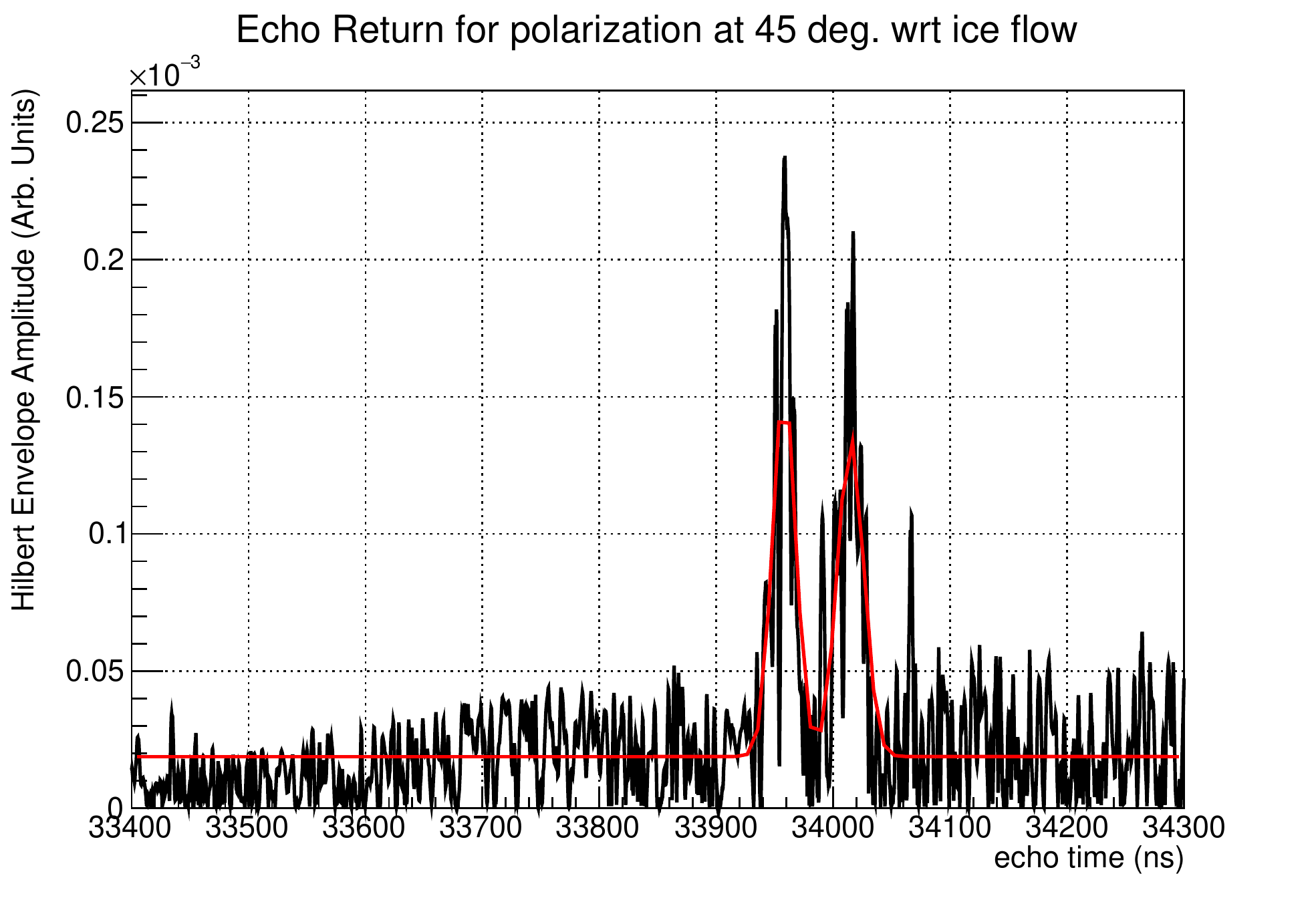}}\caption{Measured South Pole bedrock echo signal power, as function of time, for polarization at 45 degrees w.r.t. ice flow direction, showing both the fast (early) and slow (later) echo returns. Overlaid is fit to double Gaussian signal plus flat background.}\label{fig:DubPulseBedEcho.pdf}\end{figure}

Figure \ref{fig:DubPulseBedEcho.pdf} shows the bedrock return for signal polarizations at 45 degrees relative to the local ice flow direction\citep{Besson:2010ww}. Both the fast and slow bedrock echoes have comparable amplitudes, indicating that the amplitude dependence observed for internal layer scattering is a direct consequence of the scatterer itself, with a mechanism distinct from reflections at the ice/bedrock interface. 

\begin{figure}[htpb]\centerline{\includegraphics[width=0.8\textwidth]{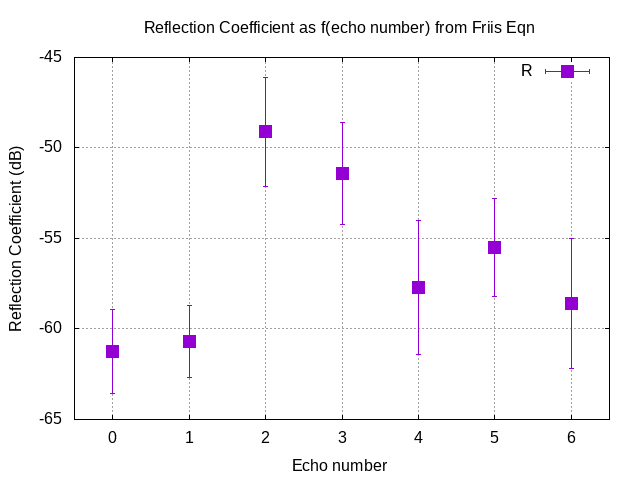}}\caption{Peak layer power reflection coefficient, calculated from radar equation\citep{shaw2013radiometry}, calculated for a frequency of 300 MHz ($\lambda$=1 m). Correlation between echo number and return time is: 0: 6 $\mu$s; 1--3: triplet between 9.67 and 9.9 $\mu$s; 4-6: triplet between 13.8 and 14.1 $\mu$s. Displayed errors are statistical only; systematic errors are estimated at $\pm$10 dB per point.}\label{fig:ReflCoeff}\end{figure}

\paragraph{Estimate of internal layer reflection strength}
Although not an original goal of the RICE echo layer measurements, we have attempted to estimate the intrinsic reflectivity of each layer. This is particularly important for current UHEN-detection experiments with estimated detection rates of order 1 per 10 years, since down-coming radio signals from either natural or anthropogenic origin may reflect upwards from an underlying embedded in-ice layer, yielding signatures otherwise indistinguishable from those of the sought-after neutrinos.
The Friis Equation\citep{shaw2013radiometry} prescribes the wavelength-dependent received power $P_{Rx}(\lambda)$ given the power output from a transmitter $P_{Tx}(\lambda)$ in the frequency band overlapping with the receiver frequency response, for propagation through a distance $d$ in a medium with frequency-dependent attenuation length $L_\alpha$:

\centerline{$P_{Rx}(\lambda)=P_{Tx}(\lambda){\cal G_{\rm Tx}}{\cal G_{\rm Rx}}f_{focus}f_{passband}(\frac{\lambda}{4\pi d})^2exp(-d/L_\alpha)$;} 

\noindent here, $\lambda$ is the wavelength being broadcast (here, we use $<\lambda>$=1 meter, following previous measurements\citep{allison2012design}), ${\cal G_{\rm Tx}}$ and ${\cal G_{\rm Tx}}$ refer to the gain of the transmitter and receiver, respectively, $f_{passband}$ prescribes the fraction of the transmitter signal power in the receiver passband, and $f_{focus}$ represents a geometric flux focusing factor. For the case of reflection, we must introduce the so-called radar cross-section (`RCS') $\sigma_{RCS}$. In that case, for bistatic measurement of a reflecting layer a distance $d$ from transmitter, we obtain:

\centerline{$P_{Rx}(\lambda)=P_{Tx}(\lambda){\cal G_{\rm Tx}}{\cal G_{\rm Rx}}f_{passband}f_{focus}F_{Fresnel}(\frac{\lambda}{4\pi d})^2exp(-2d/L_\alpha)\times (\frac{\sigma_{RCS}}{4\pi d^2})$;}

\noindent
with $F_{Fresnel}$ an angular-dependent Fresnel coefficient; this equation reduces to the Friis equation if the scattering cross-section equals $4\pi d^2$ (reflection of all the output power) and $F_{Fresnel}\equiv 1$. In our case, we take the scattering cross-section to be equal to one Fresnel zone at the distance $d$: $\sigma_{RCS}=\pi \lambda d^2/2d$, which affords an estimate of the absolute scattering power of each layer.
Results are summarized in Figure \ref{fig:ReflCoeff}. Typical peak (corresponding to $\phi\sim$150 degrees) reflection coefficients are of order --50 - --60 dB, although systematic errors are estimated to be $\pm$10 dB per point. The dominant systematic error is the uncertainty in the bandpass overlap between the transmitted signal and the received signals;\footnote{Unfortunately, the signal generator that had originally been purposed for these measurements failed in the field; we were therefore relegated to a backup unit that was less well-matched to the receiver characteristics.} the full error budget is summarized in Table \ref{tab:ReflCoeff}.
\begin{table}[htpb] 
\begin{center}
\begin{tabular}{cccc} \\
Parameter & Value & Comment & ${\cal R}$ Sys Error \\ \hhline{====}
$F_{Fresnel}$ & 1 & High conductivity reflector & 3 dB \\
$f_{focus}$ & $(1.78/1.35)^2$ & Aereal flux focusing & 2 dB \\
$L_\alpha$ & 300-1500 m & Depth-dependent & 6 dB\\
$\lambda$ & 1 m & midpoint of pass band & 4 dB \\
${\cal G_{\rm Tx}/G_{\rm Rx}}$ & 12 dB/12 dB & ANITA data\citep{besson2015antarctic} & 3 dB \\
$P_{Rx}/P_{Tx}$ & & Obtained from a fit & 6 dB \\ 
$f_{passband}$ &  0.06 & Observed power spectra & 6 dB \\ \hhline{====}
\end{tabular} 
\caption{Systematic errors in measurement of absolute reflection coefficient}
\label{tab:ReflCoeff}
\end{center}
\end{table}
Overall, our results are in general agreement with other measurements made in the polar regions (Table \ref{tab:rc}). The greatest uncertainty in our measurement is the relatively small fraction of the transmitted signal power (peaking at $\sim$100 MHz) that overlaps with the receiver passband (with a minimum frequency response of $\sim$225 MHz).


\begin{center}
\begin{table}[htpb]
\centering
\begin{tabular}{cccccc} \\ \hhline{======}
Site  & depth(m)  & $f_0$ (MHz) & pulse duration (ns) & $P_0$ (W) & ${\cal R}$ (dB) \\ \hhline{======} 
 Tuto-Camp Century\citep{bailey1964radio} (G)           & 420-610   & 35 MHz & 240 ns  & 50 W  & -43 -- -70  \\  
near Camp Century\citep{robin1969interpretation} (G)           & 410-455   & 35 MHz & 240 ns  & 50 W & -72 -- -80   \\  
Crete\citep{preben1975layer} (G)                        & 2300-2500 & 60 MHz & 250ns & 10kW  & -70 -- -75    \\  \hhline{======}

 Byrd station\citep{robin1970radio}                  & 1250      & 35MHz & 240ns & 500W   & -69  \\  
Sovetskaya\citep{robin1970radio}                  & $<$2200     & 35MHz & 240ns & 500W  & -76  \\  

South Pole\citep{harrison1973radio}  & 2000-3000 & 60MHz & 1000ns & 800W & -75 \\ \hhline{======}                                                                     
\end{tabular}
\caption{Summary of internal layer reflection coefficients measurements at variety of Greenland (designated by ``(G)'') and also Antarctic sites. Center frequency $f_0$, pulse duration and power output $P_0$ of radar system are as shown, as well as measured reflection coefficient ${\cal R}$.}
\label{tab:rc}
\end{table}
\end{center}



\begin{figure}[h]    
\centering
\includegraphics[height=0.9\textwidth,width=\textwidth]{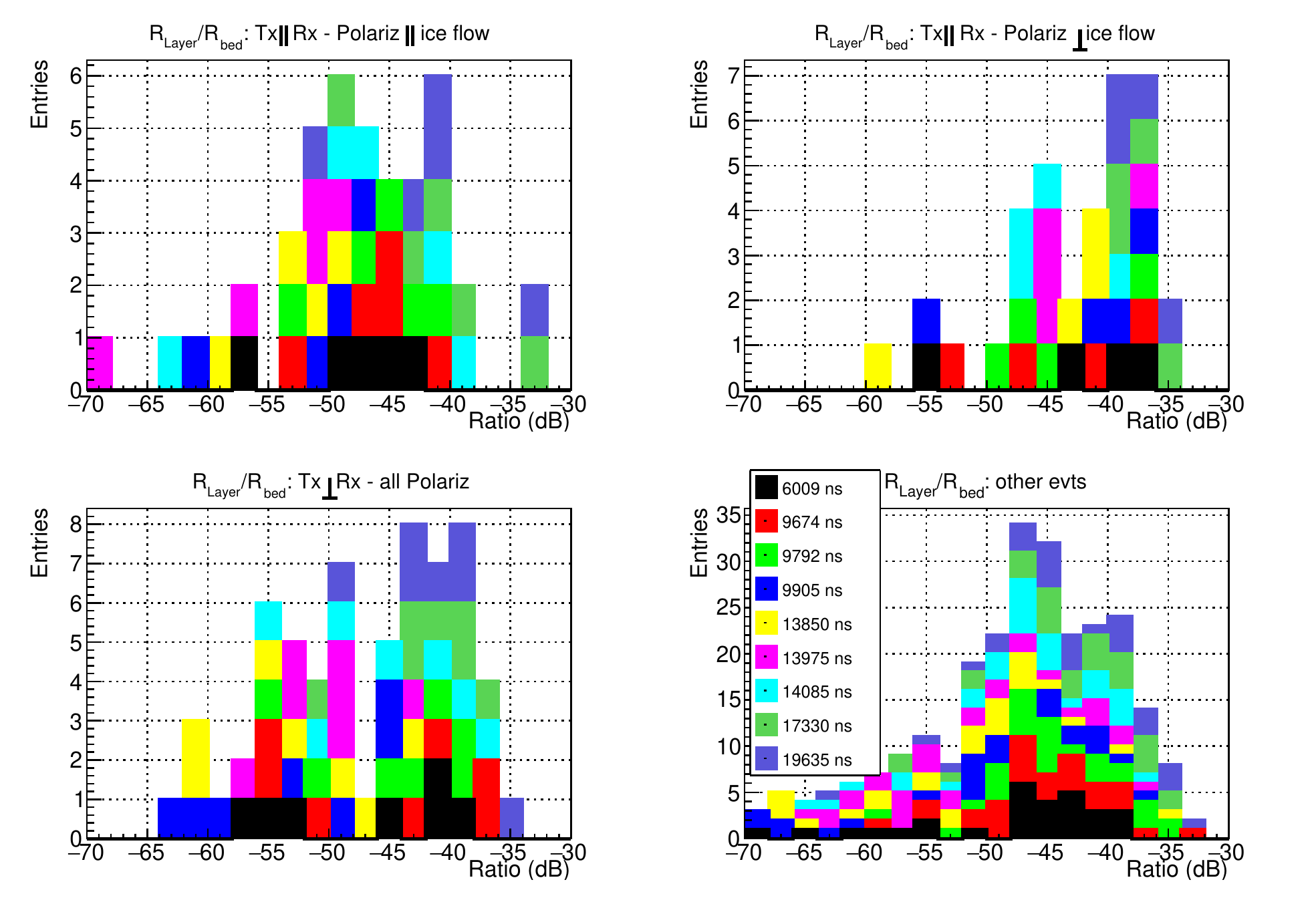} 
\caption{Reflection Coefficients measured relative to bedrock. Four different experimental configurations are presented (clockwise from top left): a) Transmitter azimuthal polarization parallel to Receiver azimuthal polarization and polarization vector parallel to the direction of the local horizontal ice flow to within 20 degrees, b) Transmitter azimuthal polarization parallel to Receiver azimuthal polarization and polarization vector perpendicular to the direction of the local horizontal ice flow to within 20 degrees, c) Transmitter azimuthal polarization perpendicular to Receiver azimuthal polarization, and d) remaining data. For each experimental configuration, relative reflectivity values have been compiled for various layers (indexed by measured echo return time) and `stacked'. Each individual echo therefore corresponds to approximately 5--10 measurements. The absolute reflectivity of each layer would be calculated by multiplying by the (unknown, but likely in the range --10$\to$--20 dB) bedrock reflectivity.}\label{fig:RLayer2Bed}\end{figure}

Alternately, for the brightest layers, we can normalize the observed peak reflectivity to the reflectivity of the bedrock. This normalization, in principle, removes the system transmitter gain and receiver response, although it does require correcting for signal attenuation in the bulk ice intervening between a given layer and the bedrock (over a two-way path). Figure \ref{fig:RLayer2Bed} presents the result of this exercise, for four different transmitter/receiver configurations sampling a range of relative transmitter/receiver orientations (either co-polarized or cross-polarized) and also sampling a range of signal polarizations (either aligned with the local ice flow direction to within $\pm 20^\circ$, transverse to the local ice flow direction to within $\pm 20^\circ$, or unaligned with either direction).  Assuming a dielectric contrast between ice ($\epsilon_{ice}\sim 3.16$) and bedrock ($\epsilon_{rock}\sim 6$) gives an expected normal incidence power reflection coefficient, for coherent specular scattering of -10 dB. Therefore, to obtain the normal incidence absolute reflectivity of any given internal layer, the distributions presented in Figure \ref{fig:RLayer2Bed} would need to be reduced by -10 dB. 
Consistent with the observed variation in reflector strength with azimuthal polarization, the bedrock-normalized layer reflectivity for polarizations perpendicular to the local ice flow direction is significantly greater than reflectivity for polarizations parallel to the local ice flow direction. These results have potential implications for estimates of backgrounds to radio neutrino detection experiments due to internally reflected radio-frequency signals originating at or near the surface.

We have further investigated the reflectivity, as a function of frequency. Dividing our sample into low- and high- bands relative to the approximately average frequency of 300 MHz, 
the peak value for the distribution is approximately 45 dB and the distributions are experimentally indistinguishable, indicating, at most, only moderate frequency dependence on reflectivity. 

\section{Summary and Conclusions}
Based on an improved analysis of vertical radar echo data taken at the South Pole in 2007 and 2009, combined with previously reported results, we summarize our conclusions as follows:
\begin{enumerate}
\item Complementing previous measurements at largely horizontal launch angles\citep{allison2020long}, we have measured birefringent asymmetries for signal propagating along ${\hat z}$ over multi-kilometer distances, reflecting from both internal layers and also the bedrock. For horizontal launch angles, the difference between the measured time-of-arrival difference asymmetry for the ARA vertically-polarized antennas relative to the ARA horizontally-polarized antennas was compared to prediction, based on the measured crystal-orientation-fabric. Parameterized as a two-variable function in terms of the opening angle $\theta$ between the signal launch direction and the local ice flow direction, and assuming that the magnitude of the asymmetry varies linearly with total signal path-length L, that previously published analysis determined that an ansatz $\delta_t$(H-V )[ns/km]=acos$\theta$+b yields values of a=8.3$\pm$1.3, and b=-8.6$\pm$0.9, with an average deviation between model and data of 12\%. For the vertical propagation case detailed herein, the data points follow prediction to within an average deviation of 14\%. If such effects can be locally calibrated, the quadrature sum of these two values indicates that, for a future South Polar radio neutrino experiment, the distance-to-vertex might be estimated to within $\sim$18\% based on birefringent asymmetries. This technique is particularly useful for distant vertices with large pathlengths through the ice. For nearby vertices, corresponding to distances of less than 0.5 km between the neutrino interaction point and the receiver array, typical resolutions of 10\% in distance can be obtained based on wavefront curvature alone -- the maximum time difference between a plane wave arrival time compared to a spherical wave arrival time from a source 500 meters away is approximately 500 picoseconds, compared to a typical cross-correlation relative timing resolution of 100 picoseconds\citep{allison2012design}. The experimentally realized distance resolution, of course, will improve with larger antenna spacing (lever arm) and the timing precision, which, in turn, will improve with higher signal-to-noise.
\item We measure the amplitude dependence of internal layer echo returns, as a function of the azimuthal polarization angle, and find an order-of-magnitude difference between the signal power returned for the case where the polarization is parallel to the ice flow direction, relative to the case where the polarization is perpendicular to the ice flow direction. This result is consistent with previous measurements made along a traverse from Dome Fuji to the Antarctic coast. 
\item Using the Friis equation, we have made an estimate of the absolute reflectivity of several internal layers, obtaining values typically ranging from --50$\to$--65 dB, but with large attendant systematic errors. This measurement is complemented by normalizing the internal layer reflectivity to the bedrock. Assuming bedrock reflection coefficients of $\approx -$10 dB yields internal layer reflectivities consistent with the `absolute' determination, using the Friis Equation.
\end{enumerate}
In order to quantify the impact of these measurements on current and planned UHEN experiments, detailed simulations of a variety of physics processes are required. Such simulations are currently underway. Refinement of these measurements would benefit from improved signal transmission, and, perhaps most importantly, higher bandwidth and sampling-rate data acquisition hardware. It is hoped that such a campaign may be conducted at the South Pole in one of the upcoming field seasons.
\section{Acknowledgments}

IK and DZB thank the National Science Foundation for their generous support of the IceCube EPSCoR Initiative (Award ID 2019597).
\bibliography{Zref}

\begin{thebibliography}{10}

\bibitem{Askaryan1962a}
G.~Askaryan.
\newblock {Excess negative charge of an electron-photon shower and its coherent
  radio emission}.
\newblock {\em Soviet Physics JETP}, 14:441--443, 1962.

\bibitem{Askaryan1965}
G.~A. {Askaryan}.
\newblock {Coherent Radio Emission from Cosmic Showers in Air and in Dense
  Media}.
\newblock {\em Soviet Phys. JETP}, 21:658, 1965.

\bibitem{ZasHalzenStanev1992}
E.~{Zas}, F.~{Halzen}, and T.~{Stanev}.
\newblock {Electromagnetic pulses from high-energy showers: Implications for
  neutrino detection}.
\newblock {\em Phys. Rev. D}, 45:362--376, January 1992.

\bibitem{Frichter:1995cn}
{Frichter, George M. and Ralston, John P. and McKay, Douglas W.}
\newblock {On radio detection of ultrahigh-energy neutrinos in Antarctic ice}.
\newblock {\em Phys. Rev.}, D53:1684--1698, 1996.

\bibitem{KravchenkoFrichterSeckel2003}
I.~{Kravchenko}, G.~M. {Frichter}, D.~{Seckel}, G.~M. {Spiczak}, J.~{Adams},
  S.~{Seunarine}, C.~{Allen}, A.~{Bean}, D.~{Besson}, D.~J. {Box}, R.~{Buniy},
  J.~{Drees}, D.~{McKay}, J.~{Meyers}, L.~{Perry}, J.~{Ralston}, S.~{Razzaque},
  and D.~W. {Schmitz}.
\newblock {Performance and simulation of the RICE detector}.
\newblock {\em Astropart. Phys.}, 19:15--36, April 2003.

\bibitem{Kravchenko:2006qc}
I.~Kravchenko et~al.
\newblock {Rice limits on the diffuse ultrahigh energy neutrino flux}.
\newblock {\em Phys. Rev.}, D73:082002, 2006.

\bibitem{SaltzbergGorhamWalz2001}
D.~{Saltzberg}, P.~{Gorham}, D.~{Walz}, and {et al.}
\newblock {Observation of the Askaryan Effect: Coherent Microwave Cherenkov
  Emission from Charge Asymmetry in High-Energy Particle Cascades}.
\newblock {\em Phys. Rev. Lett.}, 86:2802--2805, March 2001.

\bibitem{AskaryanSLAC2}
P.~W. Gorham, S.~W. Barwick, J.~J. Beatty, and {et al.}
\newblock {Observations of the Askaryan Effect in Ice}.
\newblock {\em Phys. Rev. Lett.}, 99:171101, Oct 2007.

\bibitem{barwick2005south}
{Barwick, S., Besson, D., Gorham, P. and Saltzberg, D.}
\newblock {South Polar in situ radio-frequency ice attenuation}.
\newblock {\em Journal of Glaciology}, 51(173):231--238, 2005.

\bibitem{Barwick:2014pca}
{S.W.Barwick and E.C.Berg and D.Z.Besson and G.Binder and W.R.Binns and
  D.Boersma and R.G.Bose and D.L.Braun and J.H.Buckley and V.Bugaev and
  S.Buitink and K.Dookayka and P.F.Dowkontt and T.Duffin and S.Euler and
  L.Gerhardt and L.Gustafsson and A.Hallgren and J.C.Hanson and M.H.Israel and
  J.Kiryluk and S.Klein and S.Kleinfelder and H.Niederhausen and M.A.Olevitch
  and C.Persichelli and K.Ratzlaff and B.F.Rauch and C.Reed and M.Roumi and
  A.Samanta and G.E.Simburger and T.Stezelberger and J.Tatar and U.Uggerhoj and
  J.Walker and R.Young}.
\newblock {A First Search for Cosmogenic Neutrinos with the ARIANNA Hexagonal
  Radio Array}.
\newblock {\em Astropart. Phys.}, 70:12--26, 2015.

\bibitem{Barwick:2014rca}
{S. W. Barwick and E. C. Berg and D. Z. Besson and E. Cheim and T. Duffin and
  J. C. Hanson and S. R. Klein and S. A. Kleinfelder and T. Prakash and M.
  Piasecki and K. Ratzlaff and C. Reed and M. Roumi and A. Samanta and T.
  Stezelberger and J. Tatar and J. Walker and R. Young and L. Zou}.
\newblock {Design and Performance of the ARIANNA Hexagonal Radio Array
  Systems}.
\newblock {\em IEEE Trans. Nucl. Sci.}, 62(5):2202--2215, 2015.

\bibitem{Barwick:2016mxm}
S.W. Barwick et~al.
\newblock {Radio detection of air showers with the ARIANNA experiment on the
  Ross Ice Shelf}.
\newblock {\em Astropart. Phys.}, 90:50--68, 2017.

\bibitem{Kravchenko:2011im}
Ilya Kravchenko, S.~Hussain, Dave Seckel, Dave Besson, E.~Fensholt, John
  Ralston, John Taylor, Ken Ratzlaff, and Rob Young.
\newblock {Updated Results from the RICE Experiment and Future Prospects for
  Ultra-High Energy Neutrino Detection at the South Pole}.
\newblock {\em Phys. Rev.}, D85:062004, 2012.

\bibitem{kravchenko2004situ}
Ilya Kravchenko, David Besson, and Josh Meyers.
\newblock In situ index-of-refraction measurements of the south polar firn with
  the rice detector.
\newblock {\em Journal of Glaciology}, 50(171):522--532, 2004.

\bibitem{Kravchenko:2003gk}
I.~Kravchenko, D.~Besson, J.~Drees, and J.~Meyers.
\newblock {Measurement of the radiofrequency properties of Antarctic ice with
  the RICE detector}.
\newblock In {\em {Proceedings, 28th International Cosmic Ray Conference (ICRC
  2003): Tsukuba, Japan, July 31-August 7, 2003}}, pages 1345--1348, 2003.

\bibitem{allison2012design}
P.~Allison et~al.
\newblock {Design and initial performance of the Askaryan Radio Array prototype
  EeV neutrino detector at the South Pole}.
\newblock {\em Astroparticle Physics}, 35(7):457--477, 2012.

\bibitem{allison2019design}
P~Allison, S~Archambault, R~Bard, JJ~Beatty, M~Beheler-Amass, DZ~Besson,
  M~Beydler, M~Bogdan, C-C Chen, C-H Chen, et~al.
\newblock Design and performance of an interferometric trigger array for radio
  detection of high-energy neutrinos.
\newblock {\em Nuclear Instruments and Methods in Physics Research Section A:
  Accelerators, Spectrometers, Detectors and Associated Equipment},
  930:112--125, 2019.

\bibitem{allison2020constraints}
P~Allison, S~Archambault, JJ~Beatty, M~Beheler-Amass, DZ~Besson, M~Beydler,
  CC~Chen, CH~Chen, P~Chen, BA~Clark, et~al.
\newblock Constraints on the diffuse flux of ultrahigh energy neutrinos from
  four years of askaryan radio array data in two stations.
\newblock {\em Physical Review D}, 102(4):043021, 2020.

\bibitem{Aguilar:2020xnc}
J.A. Aguilar et~al.
\newblock {Design and Sensitivity of the Radio Neutrino Observatory in
  Greenland (RNO-G)}.
\newblock 10 2020.

\bibitem{GorhamAllisonBarwick2009}
P.~W. {Gorham}, P.~{Allison}, S~W. {Barwick}, and {et al.}
\newblock {The Antarctic Impulsive Transient Antenna ultra-high energy neutrino
  detector: Design, performance, and sensitivity for the 2006-2007 balloon
  flight}.
\newblock {\em Astropart. Phys.}, 32(1):10 -- 41, 2009.

\bibitem{abarr2021payload}
Q~Abarr, P~Allison, J~Ammerman Yebra, J~Alvarez-Mu{\~n}iz, JJ~Beatty,
  DZ~Besson, P~Chen, Y~Chen, C~Xie, JM~Clem, et~al.
\newblock The payload for ultrahigh energy observations (pueo): a white paper.
\newblock {\em Journal of Instrumentation}, 16(08):P08035, 2021.

\bibitem{HooverNamGorham2010}
{S. Hoover and J. Nam and P. W. Gorham and E. Grashorn and P. Allison and S. W.
  Barwick and J. J. Beatty and K. Belov and D. Z. Besson and W. R. Binns and C.
  Chen and P. Chen and J. M. Clem and A. Connolly and P. F. Dowkontt and M. A.
  DuVernois and R. C. Field and D. Goldstein and A. G. Vieregg and C. Hast and
  C. L. Hebert and M. H. Israel and A. Javaid and J. Kowalski and J. G. Learned
  and K. M. Liewer and J. T. Link and E. Lusczek and S. Matsuno and B. C.
  Mercurio and C. Miki and P. Miočinović and C. J. Naudet and J. Ng and R. J.
  Nichol and K. Palladino and K. Reil and A. Romero-Wolf and M. Rosen and L.
  Ruckman and D. Saltzberg and D. Seckel and G. S. Varner and D. Walz and F.
  Wu}.
\newblock {Observation of Ultrahigh-Energy Cosmic Rays with the ANITA
  Balloon-Borne Radio Interferometer}.
\newblock {\em Phys. Rev. Lett.}, 105:151101, Oct 2010.

\bibitem{Prohira:2019glh}
S.~Prohira et~al.
\newblock {Observation of Radar Echoes From High-Energy Particle Cascades}.
\newblock {\em Phys. Rev. Lett.}, 124(9):091101, 2020.

\bibitem{prohira2021radar}
Steven Prohira, KD~de~Vries, Patrick Allison, J~Beatty, David Besson, Amy
  Connolly, Paramita Dasgupta, Cosmin Deaconu, S~De~Kockere, D~Frikken, et~al.
\newblock The radar echo telescope for cosmic rays: Pathfinder experiment for a
  next-generation neutrino observatory.
\newblock {\em Physical Review D}, 104(10):102006, 2021.

\bibitem{barwick2018observation}
SW~Barwick, EC~Berg, DZ~Besson, G~Gaswint, C~Glaser, Allan Hallgren, JC~Hanson,
  SR~Klein, S~Kleinfelder, L~K{\"o}pke, et~al.
\newblock Observation of classically forbidden electromagnetic wave propagation
  and implications for neutrino detection.
\newblock {\em Journal of Cosmology and Astroparticle Physics}, 2018(07):055,
  2018.

\bibitem{deaconu2018measurements}
C~Deaconu, AG~Vieregg, SA~Wissel, J~Bowen, S~Chipman, A~Gupta, C~Miki,
  RJ~Nichol, and D~Saltzberg.
\newblock Measurements and modeling of near-surface radio propagation in
  glacial ice and implications for neutrino experiments.
\newblock {\em arXiv preprint arXiv:1805.12576}, 2018.

\bibitem{telescope2021modeling}
S~Prohira, C~Sbrocco, P~Allison, J~Beatty, D~Besson, A~Connolly, P~Dasgupta,
  C~Deaconu, KD~de~Vries, et~al.
\newblock Modeling in-ice radio propagation with parabolic equation methods.
\newblock {\em Physical Review D}, 103(10):103007, 2021.

\bibitem{matsuoka2003crystal}
Kenichi Matsuoka, Teruo Furukawa, Shuji Fujita, Hideo Maeno, Seiho Uratsuka,
  Renji Naruse, and Okitsugu Watanabe.
\newblock Crystal orientation fabrics within the antarctic ice sheet revealed
  by a multipolarization plane and dual-frequency radar survey.
\newblock {\em Journal of Geophysical Research: Solid Earth}, 108(B10), 2003.

\bibitem{fujita2003scattering}
Shuji Fujita, Kenichi Matsuoka, Hideo Maeno, and Teruo Furukawa.
\newblock Scattering of vhf radio waves from within an ice sheet containing the
  vertical-girdle-type ice fabric and anisotropic reflection boundaries.
\newblock {\em Annals of Glaciology}, 37:305--316, 2003.

\bibitem{fujita2006radio}
Shuji Fujita, Hideo Maeno, and Kenichi Matsuoka.
\newblock {Radio-wave depolarization and scattering within ice sheets: a
  matrix-based model to link radar and ice-core measurements and its
  application}.
\newblock {\em Journal of Glaciology}, 52(178):407--424, 2006.

\bibitem{Kenny2012Fabric}
Kenichi Matsuoka, Donovan Power, Shuji Fujita, and Charles~F. Raymond.
\newblock {Rapid development of anisotropic ice-crystal-alignment fabrics
  inferred from englacial radar polarimetry, central West Antarctica}.
\newblock {\em {JOURNAL OF GEOPHYSICAL RESEARCH-EARTH SURFACE}}, {117}, {SEP
  18} {2012}.

\bibitem{bragg1921crystal}
WH~Bragg.
\newblock The crystal structure of ice.
\newblock {\em Proceedings of the Physical Society of London (1874-1925)},
  34(1):98, 1921.

\bibitem{rigsby1960crystal}
George~P Rigsby.
\newblock Crystal orientation in glacier and in experimentally deformed ice.
\newblock {\em Journal of Glaciology}, 3(27):589--606, 1960.

\bibitem{alley1995mapping}
Richard~B Alley, AJ~Gow, and DA~Meese.
\newblock Mapping c-axis fabrics to study physical processes in ice.
\newblock {\em Journal of Glaciology}, 41(137):197--203, 1995.

\bibitem{jordan2020modeling}
Thomas~M Jordan, DZ~Besson, I~Kravchenko, U~Latif, B~Madison, A~Nokikov, and
  A~Shultz.
\newblock Modeling ice birefringence and oblique radio wave propagation for
  neutrino detection at the south pole.
\newblock {\em Annals of Glaciology}, 61(81):84--91, 2020.

\bibitem{Voigt2017}
Donald~E. Voigt.
\newblock {c-Axis Fabric of the South Pole Ice Core, SPC14. U.S. Antarctic
  Program (USAP) Data Center. Dataset.}
\newblock 2017.

\bibitem{matsuoka1997precise}
Takeshi Matsuoka, Shuji Fujita, Shigenori Morishima, and Shinji Mae.
\newblock Precise measurement of dielectric anisotropy in ice ih at 39 ghz.
\newblock {\em Journal of Applied Physics}, 81(5):2344--2348, 1997.

\bibitem{allison2020long}
P~Allison, S~Archambault, JJ~Beatty, DZ~Besson, CC~Chen, CH~Chen, P~Chen,
  A~Christenson, BA~Clark, W~Clay, et~al.
\newblock Long-baseline horizontal radio-frequency transmission through polar
  ice.
\newblock {\em Journal of Cosmology and Astroparticle Physics}, 2020(12):009,
  2020.

\bibitem{Connolly:2021cum}
Amy Connolly.
\newblock {Impact of biaxial birefringence in polar ice at radio frequencies on
  signal polarizations in ultra-high energy neutrino detection}.
\newblock 10 2021.

\bibitem{heyer2022first}
Nils Heyer and Christian Glaser.
\newblock First-principle calculation of birefringence effects for in-ice radio
  detection of neutrinos.
\newblock {\em arXiv preprint arXiv:2205.06169}, 2022.

\bibitem{allison2019measurement}
P~Allison, S~Archambault, J~Auffenberg, R~Bard, JJ~Beatty, M~Beheler-Amass,
  DZ~Besson, M~Beydler, CC~Chen, CH~Chen, et~al.
\newblock Measurement of the real dielectric permittivity $\epsilon_r$ of
  glacial ice.
\newblock {\em Astroparticle Physics}, 108:63--73, 2019.

\bibitem{Besson:2009zza}
D.~Besson, R.~Keast, and R.~Velasco.
\newblock {In situ and laboratory studies of radiofrequency propagation through
  ice and implications for siting a large-scale Antarctic neutrino detector}.
\newblock {\em Astropart. Phys.}, 31:348--358, 2009.

\bibitem{Besson:2010ww}
Dave Besson, Ilya Kravchenko, Andres Ramos, and Juliet Remmers.
\newblock {Radio Frequency Birefringence in South Polar Ice and Implications
  for Neutrino Reconstruction}.
\newblock {\em Astropart. Phys.}, 34:755--768, 2011.

\bibitem{besson2008situ}
DZ~Besson, J~Jenkins, S~Matsuno, J~Nam, M~Smith, SW~Barwick, JJ~Beatty,
  WR~Binns, C~Chen, P~Chen, et~al.
\newblock In situ radioglaciological measurements near taylor dome, antarctica
  and implications for ultra-high energy (uhe) neutrino astronomy.
\newblock {\em Astroparticle Physics}, 29(2):130--157, 2008.

\bibitem{kennedy2013evolution}
Joseph~H Kennedy, Erin~C Pettit, and Carlos~L Di~Prinzio.
\newblock The evolution of crystal fabric in ice sheets and its link to climate
  history.
\newblock {\em Journal of Glaciology}, 59(214):357--373, 2013.

\bibitem{dzbilya2013cryo}
D.~Besson and I.~Kravchenko.
\newblock {Radio-frequency probes of Antarctic ice at South Pole}.
\newblock {\em {CRYOSPHERE}}, {7}({3}):{855--866}, {2013}.

\bibitem{shaw2013radiometry}
Joseph~A Shaw.
\newblock Radiometry and the friis transmission equation.
\newblock {\em American journal of physics}, 81(1):33--37, 2013.

\bibitem{besson2015antarctic}
DZ~Besson, J~Stockham, M~Sullivan, P~Allison, JJ~Beatty, K~Belov, WR~Binns,
  C~Chen, P~Chen, JM~Clem, et~al.
\newblock {Antarctic radio frequency albedo and implications for cosmic ray
  reconstruction}.
\newblock {\em Radio Science}, 50(1):1--17, 2015.

\bibitem{bailey1964radio}
JT~Bailey, S~Evans, and G~de~Q Robin.
\newblock Radio echo sounding of polar ice sheets.
\newblock {\em Nature}, 204(4957):420--421, 1964.

\bibitem{robin1969interpretation}
G~de~Q Robin, S~Evans, and Jeremy~T Bailey.
\newblock Interpretation of radio echo sounding in polar ice sheets.
\newblock {\em Philosophical Transactions of the Royal Society of London.
  Series A, Mathematical and Physical Sciences}, 265(1166):437--505, 1969.

\bibitem{preben1975layer}
Gudmandsen Preben.
\newblock Layer echoes in polar ice sheets.
\newblock {\em Journal of Glaciology}, 15(73):95--101, 1975.

\bibitem{robin1970radio}
G~de~Q Robin, CWM Swithinbank, BME Smith, et~al.
\newblock Radio echo exploration of the antarctic ice sheet.
\newblock {\em International Association of Scientific Hydrology Publication},
  86:97--115, 1970.

\bibitem{harrison1973radio}
CH~Harrison.
\newblock Radio echo sounding of horizontal layers in ice.
\newblock {\em Journal of glaciology}, 12(66):383--397, 1973.

\end{thebibliography}
\bibliographystyle{unsrt} 

\end{document}